\documentstyle[prl,aps,multicol,epsf]{revtex}

\newcommand{\placefig}[3]
{
 \begin{figure}[b]
 \begin{center}
  \leavevmode
  \epsfxsize=#2
  \epsfbox{#1.ps}
 \end{center}
  \caption{\label{#1} \narrowtext \sl #3}
 \end{figure}
}

\begin{document}

\title{Absence of non-trivial asymptotic scaling in the \\
Kashchiev model of polynuclear growth} 
\author{T.J. Newman and A. Volmer} 
\address{Institut f\"ur Theoretische Physik,
  Universit\"at zu K\"oln, D-50937 K\"oln, Germany \\}
\date{\today} 
\maketitle
\begin{abstract}
  In this brief comment we show that, contrary to previous claims
  [Bartelt M C and Evans J W 1993 {\it J.\ Phys.\ A} ${\bf 26}$ 2743],
  the asymptotic behaviour of the Kashchiev model of polynuclear
  growth is trivial in all spatial dimensions, and therefore lies
  outside the Kardar-Parisi-Zhang universality class.
\end{abstract}

\begin{multicols}{2}

Within the field of non-equilibrium interface growth, one of the
central issues is the existence of dynamical scaling and associated
universality classes\cite{rev1}\cite{rev2}. The simplest quantity
which may exhibit such scaling is the interface width $w(L,t)$.  For a
system of lateral size $L$, the scaling form for $w$ may be written as
$w(t) \sim L^{\chi}f(t/L^{z})$ for times $t$ larger than some
microscopic time scale. In the limit of infinite $L$, one then expects
the asymptotic time evolution of the width to follow $w \sim
t^{\beta}$ where $\beta = \chi/z$. The determination of the exponents
$z$ and $\chi $ represents a primary objective within this field.  One
of the most popular models of interface growth is due to Kardar,
Parisi and Zhang (KPZ) \cite{kpz}, and through a concerted effort
(mostly on the numerical front), there now exist rather precise
estimates for the exponents in low spatial dimensions (see for example
\cite{num}.)  An analytic derivation of these exponents represents an
outstanding theoretical challenge.

In a recent paper\cite{be}, Bartelt and Evans (BE) presented an
analysis of the Kashchiev model\cite{kash}, which is closely related to
polynuclear growth (PNG) models\cite{png}\cite{kw}\cite{ks}.
This model has the unusual
feature of being exactly solvable in the sense that the average
interface height $h(t)$ and the width $w(t)$ may be expressed in terms
of a closed set of coupled integral equations. A numerical study of
these equations was undertaken by BE. From their results, they
concluded that the interface width of the Kashchiev model showed
non-trivial asymptotic scaling with exponents which were consistent
(within numerical error) with KPZ universality. They also raised the
possibility that an exact asymptotic analysis of this model may enable
one to determine the upper critical dimension (above
which $\beta =0$) of the KPZ model.

To investigate such a possibility was our motivation for a closer
analysis of the Kashchiev model. Our results (to be presented below)
convincingly demonstrate that the asymptotic behaviour of the
Kashchiev model is trivial in the sense that the growth exponent
$\beta (d) = 1/2$ for all $d$.  This model therefore lies outside the
KPZ universality class. Our
presentation shall henceforth be brief, and we refer the reader to
ref.\cite{be} for details of the formulation of the Kashchiev model.

The central quantity in the Kashchiev model is the function $\theta
_{i}(t)$ which represents the fraction of layer $i$ which is covered
with deposited material at time $t$. The form of $\theta _{1}(t)$ is
known exactly, and the higher functions $\theta _{i > 1}(t)$ may
be generated iteratively via the integral equations
\begin{equation}
\label{e1}
\theta _{i+1}(t) = \int \limits _{0}^{t} dt' \ \left \lbrace 1 - \exp
  [-(t-t')^{d+1}] \right \rbrace {d\theta_{i}(t') \over dt' }.
\end{equation}
In terms of these layer coverages $\theta _{i}$, the average height may
be expressed as
\begin{equation}
\label{e2}
h(t) = \sum \limits _{i=1}^{\infty} \theta _{i}(t),
\end{equation}
and the mean square width $W(t) \equiv w(t)^{2}$ takes the form
\begin{equation}
\label{e3}
W(t) = \sum \limits _{i=1}^{\infty} (2i-1)\theta _{i}(t) - h(t)^2.
\end{equation}
(These expressions differ in a minor, inessential way from those of BE.
Note also that our unit of time is exactly half that used by BE.)

The numerical analysis of BE seems to have proceeded by iteratively
solving the set of integral equations for a given number of the
functions $\theta _{i}(t)$, and then summing these functions in order
to determine $h(t)$ and $W(t)$. This method will fail for large times
as one is forced to calculate an increasingly large number of layer
coverages to ensure numerical precision.  The predictions of
asymptotic scaling in accord with KPZ scaling was made on the basis of
calculating the first five layer coverages, which is 
insufficient to investigate the true asymptotic regime.

A simple way around this problem is to sum the recursion relation
(\ref{e1}) over {\it all} the layer coverages $\theta _{i}$ with an
appropriate weight such that one derives closed integral equations for
$h(t)$ and $W(t)$. One then has no numerical barriers in probing the
deep asymptotic regime.  The equation for the mean height, written in
terms of the deviation from linear growth $\Delta (t) \equiv h(t) -
v(d)t$, takes the form
\begin{samepage}
\begin{eqnarray}
\label{e4}
&&\int \limits _{0}^{t} dt' \ {d\Delta (t')\over dt'} \ 
\exp [-(t-t')^{d+1}] = \\
&&\quad 1 - \exp (-t^{d+1}) - v(d)\int \limits _{0}^{t} dt \
\exp (-t'^{d+1}). \nonumber
\end{eqnarray}
\end{samepage}
The $d$-dependent velocity
is given by $v(d)^{-1} = \Gamma \left ((d+2)/(d+1)\right )$, 
where $\Gamma (z)$ is
the gamma function\cite{as}.  For the mean square width, we have
\begin{eqnarray}
\label{e5}
&& \int \limits _{0}^{t} dt' \ {dW (t')\over dt'} \ 
\exp [-(t-t')^{d+1}]= \\
&& \quad 2h(t) - \int \limits _{0}^{t} dt' 
\ {dh(t')\over dt'} \ [1+2h(t')] \ 
\exp [-(t-t')^{d+1}] .\nonumber
\end{eqnarray}
The above equations are of the Volterra type, and we may therefore use
relatively simple techniques for their numerical solution.  More
precisely, we use a uniform grid for the discretization, along with a
trapezoidal rule for the integration\cite{nr}. The discrete time step
used here is $\delta t = 0.001$, yielding results with a precision of
6 significant figures. We should emphasize that on reducing the time
step further, the precision can be systematically improved.

The reason for studying the deviation from linear growth of $h(t)$,
via the function $\Delta (t)$, is that one might expect on simple
scaling grounds that $\Delta (t) \sim t^{\beta}$.  In Fig.~1 we
present our results for $d\Delta (t)/dt$, for $d=1,2$ and $3$. The
function becomes increasingly oscillatory for higher dimensions. The
decay of the envelope of the oscillations is exponential after some
transient period (which grows slowly with increasing dimension.) We
have determined the decay rate $\lambda (d)$ for the exponential decay
with high precision as a function of $d$. The inset in Fig.~1 shows
the $d$-dependence of $\lambda $ on a log-linear scale. In Fig.~2, we
show the evolution of the time derivative of $W(t)$ for $d=1,2$ and
$3$.  Results for higher dimensions are qualitatively similar. Again
we see that the pre-asymptotic behaviour is increasingly oscillatory
as one increases the dimension $d$.  The asymptotic behaviour,
however, is purely constant, implying that $W(t) \sim b(d) \ t$, and
hence that $\beta = 1/2$ for all $d$. The full dimensional dependence
of $b(d)$ is plotted in Fig.~3. One may also study the pre-asymptotic
corrections to this linear form for $W(t)$. These oscillatory
corrections are also found to have an exponentially decaying envelope,
with a decay rate ${\tilde \lambda }(d)$ which satisfies the relation
$\lambda (d) \simeq 1.14(2) \ {\tilde \lambda}(d)$.  These results
convincingly show that the asymptotic properties of the Kashchiev
model are trivial; i.e. the mean height relaxes exponentially fast to
linear growth, and the interface width evolves as the square root of
time.

\placefig{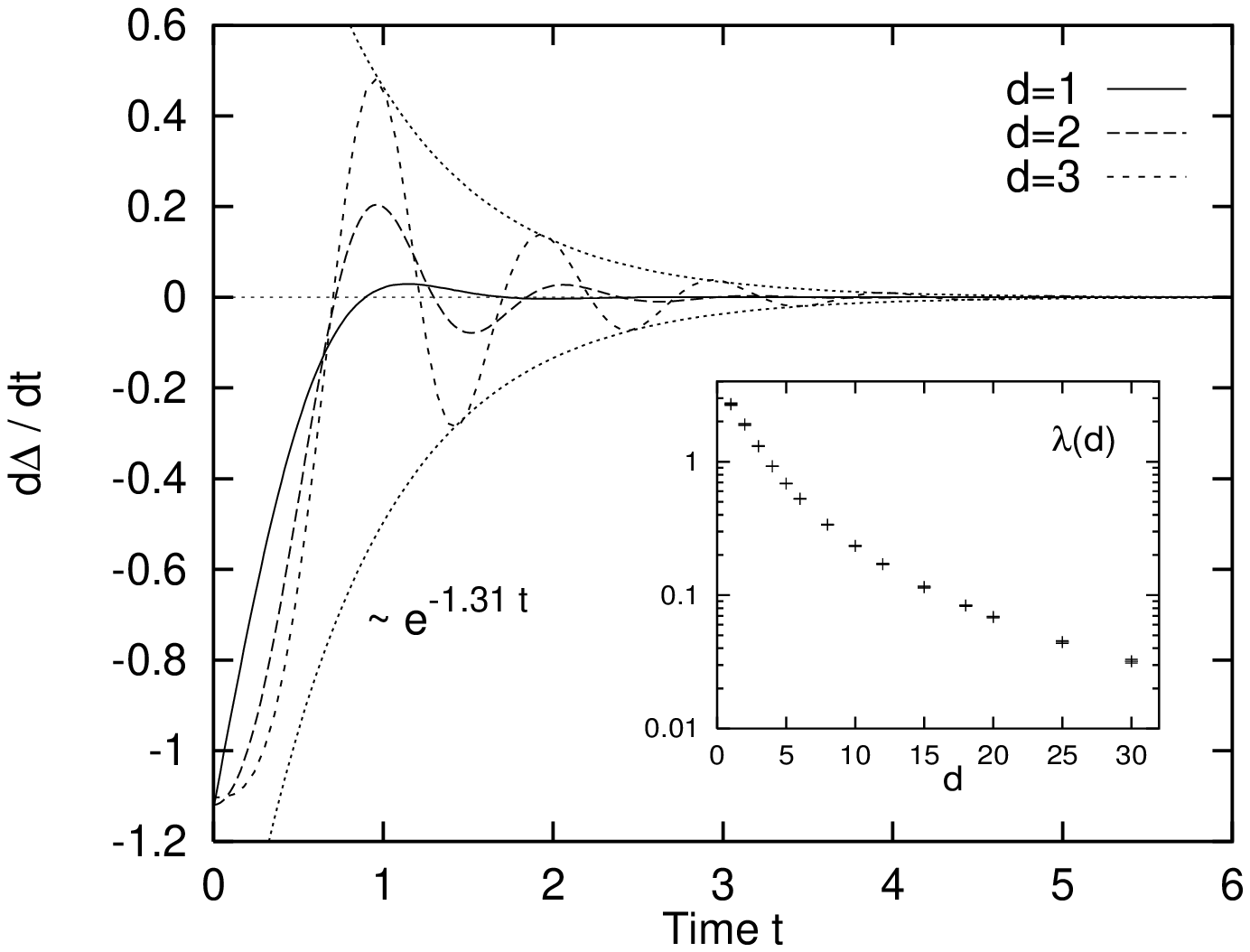}{7cm}{The time derivative of the deviation of $h(t)$ from
linear growth, $d\Delta(t)/dt$, as a function of $t$, for
$d=1,2,3$. This data comes from numerical integration of
Eq.~(\ref{e4}). The exponential decay of the envelope of oscillations
is illustrated for the $d=3$ curve.  The inset shows the
$d$-dependence of the decay rate $\lambda $ on a log-linear scale.}

\placefig{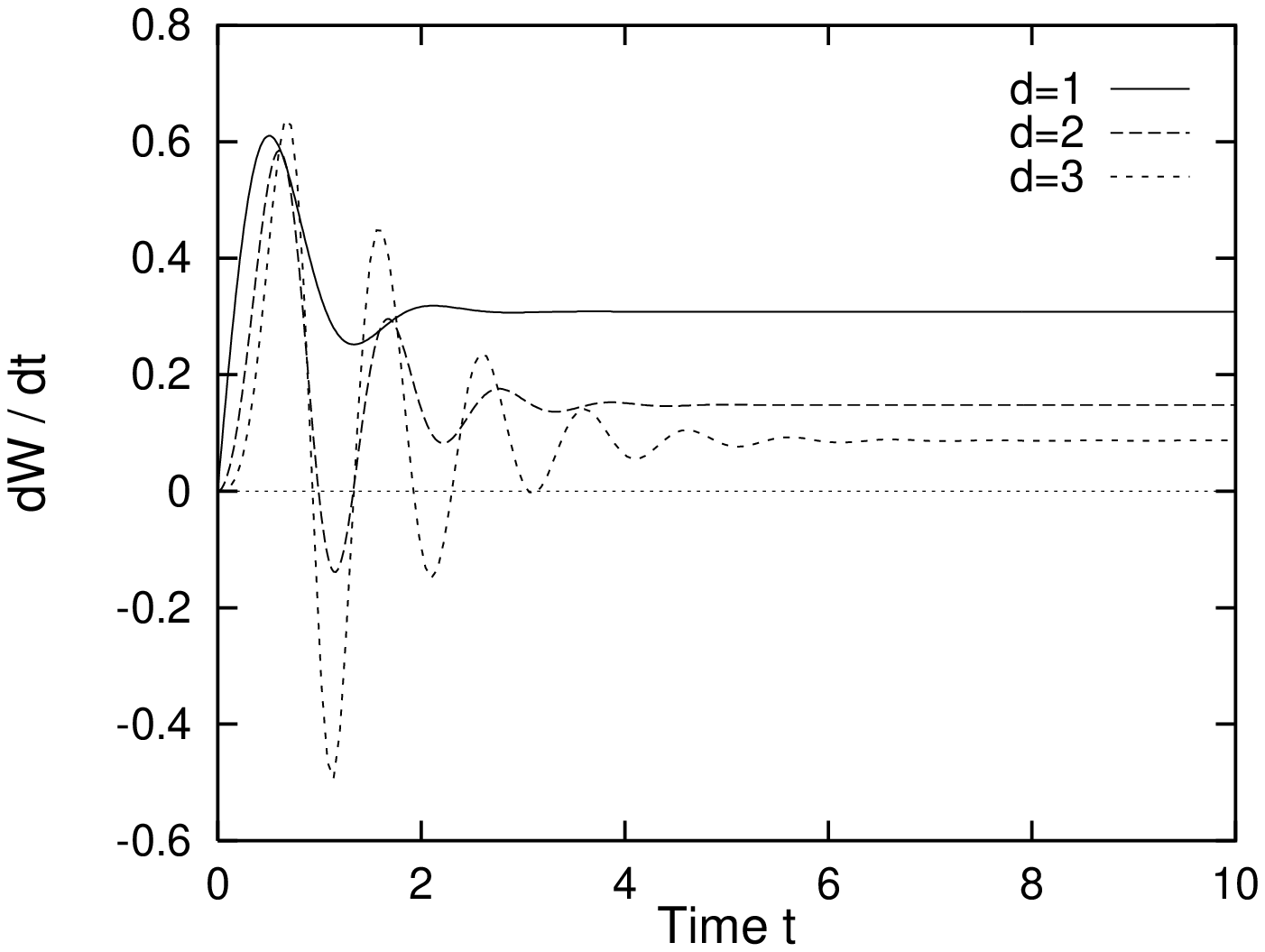}{7cm}{The time derivative of the mean square
fluctuations, $dW(t)/dt$, as a function of $t$, for $d=1,2,3$. This
data comes from simultaneous numerical integration of Eq.~(\ref{e4})
and (\ref{e5}).}

\placefig{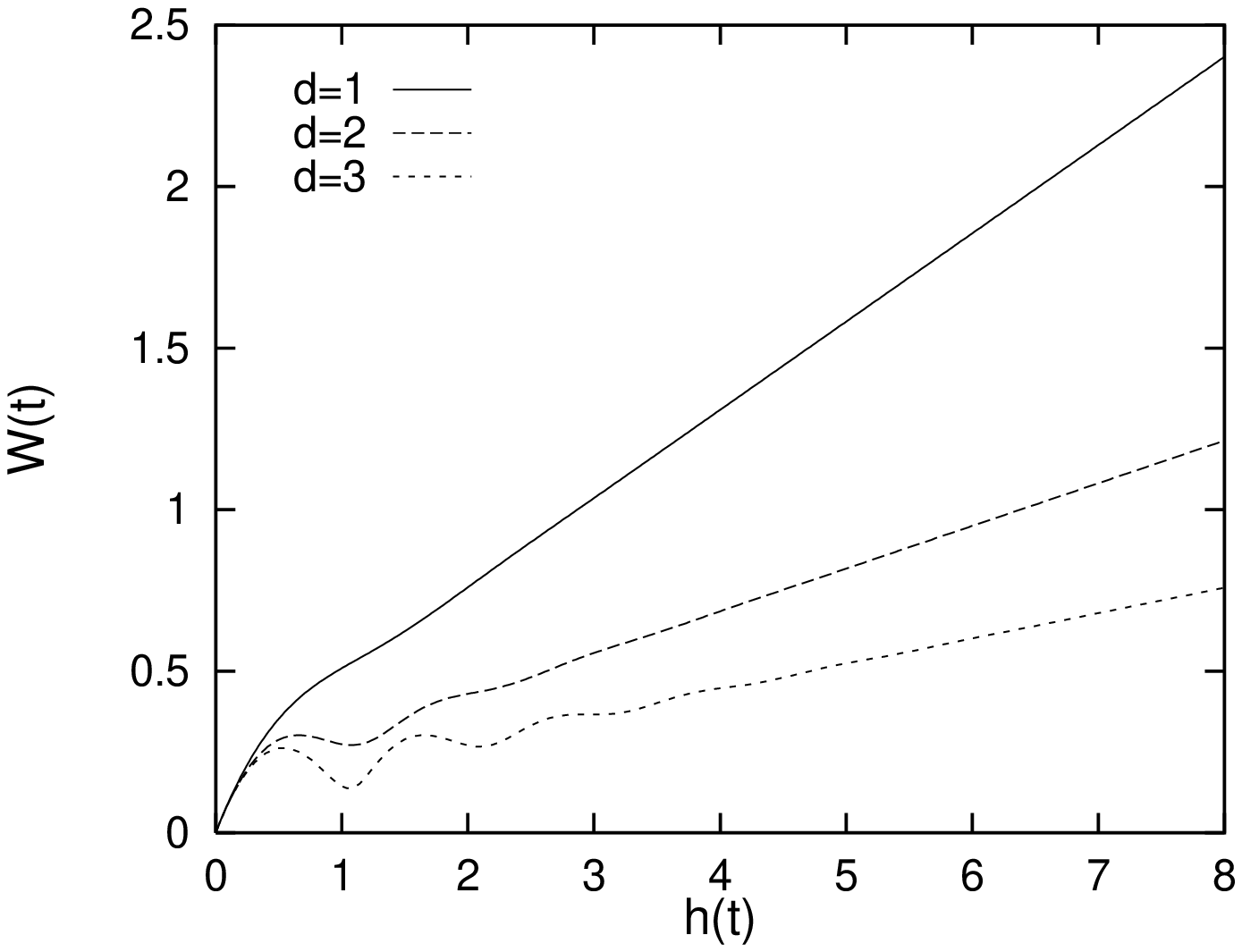}{7cm}{The $d$ dependence of the amplitude $b(d)$, from
the asymptotic relation $dW(t)/dt \sim b(d)$. The solid line
is the relation given in Eq.~(\ref{e6}), whilst the data points
are from numerical integration of Eq.~(\ref{e5}).}

As a powerful check on the numerical work, we may solve the integral
equation for the mean square width $W(t)$ in the deep asymptotic
regime by making the Ansatz $dW(t)/dt \sim b(d)$, and taking $h(t)
\sim v(d)t$, with $v(d)$ defined previously. Inserting these forms
into Eq.~(\ref{e5}) and taking $t \rightarrow \infty$ allows one to
extract the prefactor $b(d)$. Explicitly one finds
\begin{equation}
\label{e6}
b(d) = v(d) \left \lbrace \Gamma \left ( { d+3\over d+1} \right ) \ 
  v(d)^{2} - 1 \right \rbrace .
\end{equation}
This functional form is plotted against the numerical results in
Fig.~3. Excellent agreement is obtained, indicating both the
correctness of the above Ansatz, and the high precision of the
numerical work.

For comparison with the above results, we mention that the above
equations may be easily solved in the limits of $d=0$ and $d=\infty$
by use of Laplace transform methods. One finds for $d=0$, that
$h(t)=t$ and $W(t)=t$ exactly. For $d=\infty$ one has layer-by-layer
growth described by $h(t)=n, \ n < t < n+1$ and $W(t)=0$ exactly. The
infinite $d$ results are consistent with i) the increasingly
oscillatory behaviour found numerically for larger values of $d$ (a
trend towards layer-by-layer growth), and with ii) $b(d)$ vanishing
monotonically for large $d$.

Finally we illustrate the misinterpretation of the data that led BE to
conclude that $W(t)$ had KPZ-type scaling. Their conclusions arose
from a plot of $W(t)$ versus $h(t)$. Since their data is not in the
asymptotic regime, such a plot is potentially misleading as $h(t)$
still has appreciable deviations from linear growth. In Fig.~4, we
present a linear plot of $W(t)$ versus $h(t)$ for $d=1,2$ and $3$ for
longer times than obtained by BE (cf. Fig.~4 in \cite{be}.) It is seen
that the relationship between $W(t)$ and $h(t)$ eventually becomes
linear, in agreement with the scaling forms $h(t) \sim v(d)t$ and
$W(t) \sim b(d)t$.

We conclude by remarking that models belonging to the PNG class are
certainly worthy of study, as more realistic versions are known to
exhibit KPZ scaling\cite{kw}\cite{ks}. However, any analytic treatment
must be based on a less mean-field-like formulation than that of the
Kashchiev model. It is of interest to determine precisely where the
crucial mean-field assumption enters into the Kashchiev model, and
whether it is possible to relax this assumption without making the
model analytically intractable.

\placefig{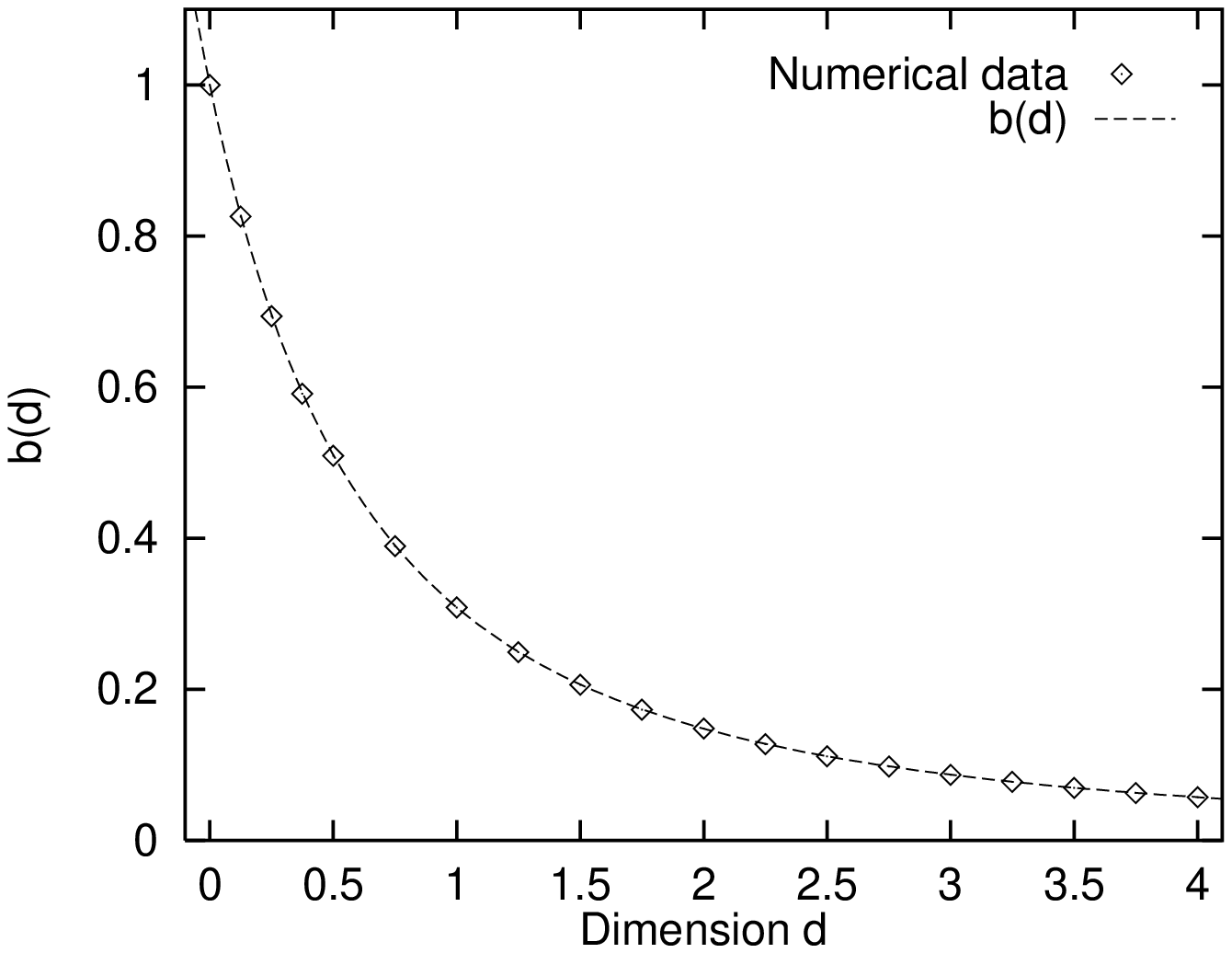}{7cm}{A plot of the mean square fluctuations $W(t)$
versus the average height $h(t)$, for $d=1,2,3$.}

It is a pleasure to thank our colleague Lei-Han Tang for interesting
discussions. The authors acknowledge financial support under grant
SFB 341 (B8).

\end{multicols}
\end{document}